\documentclass{aip-cp}

\usepackage{etoolbox}
\makeatletter
\patchcmd{\@@tablenote}{\xdef}{\protected@xdef}{}{}
\makeatother

\usepackage[numbers]{natbib}
\usepackage{rotating}
\usepackage{graphicx}

\graphicspath{{art/}}

\begin{document}

\title{Time-dependent coupled harmonic oscillators: Comment}

\author[aff1]{H.M. Moya-Cessa}
\author[aff2]{J. R\'ecamier}

\affil[aff1]{Instituto Nacional de
 Astrof\'{\i}sica, \'Optica y Electr\'onica, Calle Luis Enrique Erro
  No. 1, Santa Mar\'{\i}a Tonanzintla, Puebla, 72840, Mexico.}
\affil[aff2]{Instituto de Ciencias F\'{\i}sicas, Universidad
 Nacional Aut\'onoma de M\'exico, Apdo. Postal 48-3, Cuernavaca,
  Morelos 62251, Mexico.}

\maketitle

\begin{abstract}
Macedo and Guedes  showed recently how to solve a system of coupled harmonic oscillators with time dependent parameters [{ J. Math. Phys.} {\bf 53},  052101 (2012)]. We show here that the way in which they get rid of the time dependent masses is incorrect and some terms lack in the transformed Hamiltonian. We also show a correct way of eliminating from the Hamiltonian the time dependent masses.
\end{abstract}


Macedo and Guedes \cite{Macedo2012} have studied the interaction of two coupled time dependent oscillators \cite{Macedo2012} and have used a time dependent transformation that takes the two different time dependent masses of the single oscillators to an effective single time dependent mass. We show here that what they do is wrong and show a correct form of eliminating the masses. 

First, consider the  Hamiltonian for a single time dependent harmonic oscillator \cite{Guasti2003,JPA,Ramos2018}
\begin{equation}
 {H}=\frac{ {P}^2}{2M(t)}+\frac{M(t) {X}^2}{2},
\end{equation}
with constant frequency equal to one. Following Macedo and Guedes \cite{Macedo2012,Macedo2014} we could do $x=M(t) X$ and $p=\frac{1}{M(t)}P$, such that we would obtain a Hamiltonian that looks time independent
\begin{equation}
 {H}=\frac{ {p}^2}{2}+\frac{ {x}^2}{2},
\end{equation}
but is time dependent since the new position  and momentum operators are explicitly time dependent. In fact, eliminating the time dependent mass is not a an simple task, as we show here for the case of two coupled time  dependent oscillators Hamiltonian.

Given the Hamiltonian \cite{Macedo2012,Macedo2014}
\begin{equation}
 {H}(t)=\frac{ {p}_{1x}^2}{2m_1(t)}+\frac{ {p}_{2x}^2}{2m_2(t)}+\frac{m_1(t)\omega_1^2(t) {x}_1^2}{2}+\frac{m_2(t)\omega_2^2(t) {x}_2^2}{2}+\frac{k(t)( {x}_2- {x}_1)^2}{2},
\end{equation}
and the equation
\begin{equation}
i\frac{\partial |\psi(t)\rangle}{\partial t}= {H}(t)|\psi(t)\rangle
\end{equation}
by doing the  squeezing \cite{Yuen,Caves,Vidiella} unitary transformation $T_u|\psi(t)\rangle=|\phi(t)\rangle$, with
\begin{equation}
 {T}_{u}=e^{i\frac{ u_1(t)}{2 }( {x}_1 
{p}_{1x}+ {p}_{1x} {x}_1)}e^{i\frac{ u_2(t)}{2 }( {x}_2 {p}_{2x}+ {p}_{2x} {x}_2)}
\end{equation}
that produces
\begin{equation}
 {T}_{u} {x}_1 {T}^{\dagger}_{u}= {x}_1e^{u_1(t)}, \qquad  {T}_{u} {x}_2 {T}^{\dagger}_{u}= {x}_2e^{u_2(t)}
\end{equation}
\begin{equation}
 {T}_{u} {p}_{1x} {T}^{\dagger}_{u}= {p}_{1x}e^{-u_1(t)}, \qquad  {T}_{u} {p}_{2x} {T}^{\dagger}_{u}= {p}_{2x}e^{-u_2(t)},
\end{equation}
we obtain
\begin{equation}
i\frac{\partial  {T}^{\dagger}_{u}}{\partial t}|\phi(t)\rangle+i {T}^{\dagger}_{u}\frac{\partial |\phi(t)\rangle}{\partial t}= {H}(t) {T}^{\dagger}_{u}|\phi(t)\rangle.
\end{equation}
with
\begin{equation}
\frac{\partial  {T}^{\dagger}_{u}}{\partial t}=-i {T}^{\dagger}_{u}[\frac{\dot{u}_1}{2}( {x}_1 
{p}_{1x}+ {p}_{1x} {x}_1)+\frac{\dot{u}_2}{2}( {x}_2 
{p}_{2x}+ {p}_{2x} {x}_2)],
\end{equation}
that, by substituting in the equation above and multiplying by  $ {T}_u$ by the left gives
\begin{eqnarray}
i\frac{\partial |\phi(t)\rangle}{\partial t}+[\frac{\dot{u}_1}{2}( {x}_1 
{p}_{1x}+ {p}_{1x} {x}_1)+\frac{\dot{u}_2}{2}( {x}_2 
{p}_{2x}+ {p}_{2x} {x}_2)]|\phi(t)\rangle= {T}_u {H}(t) {T}^{\dagger}_{u}|\phi(t)\rangle.
\end{eqnarray}
or
\begin{eqnarray}
i\frac{\partial |\phi(t)\rangle}{\partial t}= {\tilde{H}}(t)|\phi(t)\rangle.\label{phi}
\end{eqnarray}
with
\begin{eqnarray}
 {\tilde{H}}(t)&=&-\frac{\dot{u}_1}{2}( {x}_1 
{p}_{1x}+ {p}_{1x} {x}_1)-\frac{\dot{u}_2}{2}( {x}_2 
{p}_{2x}+ {p}_{2x} {x}_2)\\ \nonumber&+&
\frac{ {p}_{1x}^2e^{-2u_1(t)}}{2m_1(t)}+\frac{ {p}_{2x}^2e^{-2u_2(t)}}{2m_2(t)}+\frac{m_1(t)\omega_1^2(t) {x}_1^2e^{2u_1(t)}}{2}+\frac{m_2(t)\omega_2^2(t) {x}_2^2e^{2u_2(t)}}{2}\\ \nonumber &+&\frac{k(t)( {x}_2e^{u_2(t)}- {x}_1e^{u_1(t)})^2}{2}.
\end{eqnarray}
With the choice $u_j=\ln \sqrt{\frac{m(t)}{m_j(t)}}$ we obtain the first part of the Hamiltonian (6) in Macedo and Guedes. However, note that the first term on the r.h.s. of equation (10) is lacking in their equation (6). The parameter $m(t)$ is an arbitrary function of time that we could set as $m(t)=\sqrt{m_1(t)m_2(t)}$ to obtain a result more similar to the one obtained by Macedo and Guedes. Here we prefer to set it to one, i.e., $u_j=-\frac{1}{2}\ln {m_j}$, then we get
\begin{eqnarray}
 {\tilde{H}}(t)&=&-\frac{\dot{u}_1}{2}( {x}_1 
{p}_{1x}+ {p}_{1x} {x}_1)-\frac{\dot{u}_2}{2}( {x}_2 
{p}_{2x}+ {p}_{2x} {x}_2)\\ \nonumber&+&
\frac{ {p}_{1x}^2}{2}+\frac{ {p}_{2x}^2}{2}+\frac{\omega_1^2(t) {x}_1^2}{2}+\frac{\omega_2^2(t) {x}_2^2}{2}+\frac{k(t)( {x}_2e^{u_2(t)}- {x}_1e^{u_1(t)})^2}{2},
\end{eqnarray}
that may be rewritten to give
\begin{eqnarray}
 {\tilde{H}}(t)=
\frac{ ({p}_{1x}-\frac{\dot{u}_1}{2}x_1)^2}{2}+\frac{ ({p}_{2x}-\frac{\dot{u}_2}{2}x_2)^2}{2}+\frac{(\omega_1^2-\frac{\dot{u}_1^2}{4}) {x}_1^2}{2}+\frac{(\omega_2^2-\frac{\dot{u}_2^2}{4}) {x}_2^2}{2}+\frac{k(t)( {x}_2e^{u_2}- {x}_1e^{u_1})^2}{2}.
\end{eqnarray}
By doing the transformation $R|\phi(t)\rangle=|\xi(t)\rangle$, with $R=\exp\{-i(\frac{\dot{u}_1}{4}x_1^2+\frac{\dot{u}_2}{4}x_2^2)\}$ and, inserting it into equation (\ref{phi}), we obtain
\begin{eqnarray}
i\frac{\partial (R^{\dagger} |\xi(t)\rangle)}{\partial t}= {\tilde{H}}(t)R^{\dagger}|\xi(t)\rangle,\label{xi}
\end{eqnarray}
that is rewritten as
\begin{eqnarray}
i\frac{\partial  |\xi(t)\rangle}{\partial t}=\left[\frac{\ddot{u}_1}{4}x_1^2+\frac{\ddot{u}_2}{4}x_2^2+ R{\tilde{H}}(t)R^{\dagger}\right]|\xi(t)\rangle,\label{xi2}
\end{eqnarray}
that finally yields
\begin{eqnarray}
i\frac{\partial  |\xi(t)\rangle}{\partial t}=\frac{1}{2}\left[{ {p}_{1x}^2}+{ {p}_{2x}^2}+{(\omega_1^2-\frac{\dot{u}_1^2-2\ddot{u}_1}{4}) {x}_1^2}+{(\omega_2^2-\frac{\dot{u}_2^2-2\ddot{u}_2}{4}) {x}_2^2}+{k(t)( {x}_2e^{u_2}- {x}_1e^{u_1})^2}\right]|\xi(t)\rangle.\label{xi3}
\end{eqnarray}
In Ref. \cite{Urzua} it has been  shown how this equation may be solved for time dependent arbitrary parameters.

In conclusion, we have shown how to eliminate the time dependent masses from the two coupled, time dependent, harmonic oscillators Hamiltonian.

\end{document}